\documentclass[prd,noshowpacs,nofootinbib]{revtex4}

\usepackage{color} 
\usepackage{amssymb}
\usepackage{amsmath}
\usepackage{graphicx} 
\usepackage{float}
\usepackage{braket}     
\usepackage{pdfpages}
\usepackage{epstopdf}
\usepackage[utf8]{inputenc}
\usepackage{comment}
\usepackage{xspace}
\newcommand{\hX}{\hat{X}}
\newcommand{\hP}{\hat{P}}
\newcommand{\hx}{\hat{x}}
\newcommand{\hp}{\hat{p}}
\newcommand\ie{\mbox{\textit{i.\,e.}}\xspace}
\newcommand\eg{\mbox{\textit{e.\,g.}}\xspace}
\DeclareUnicodeCharacter{202F}{\,}

\begin{document}
 
\title{Momentum gauge fields from curved momentum space through Kaluza-Klein reduction}

\author{Eduardo Guendelman}
\email{guendel@bgu.ac.il}  \email{guendelman@fias.uni-frankfurt.de}
\affiliation{Department of Physics, Ben-Gurion University of the Negev, Beer-Sheva, Israel \vspace{0.15in} \\
Frankfurt Institute for Advanced Studies, Giersch Science Center, Campus Riedberg, Frankfurt am Main, Germany \vspace{0.15in} \\
Bahamas Advanced Studies Institute and Conferences, 4A Ocean Heights, Hill View Circle, Stella Maris, Long Island, The Bahamas}

\author{ Fabian Wagner}
\email{fabian.wagner@usz.edu.pl}
\affiliation{University of Szczecin, Department of Physics, Wielkopolska 15, PL-71-061 Szczecin, Poland\\
Dipartimento di Ingegneria Industriale, Universit\`a degli Studi di Salerno, Via Giovanni Paolo II, 132 I-84084 Fisciano (SA), Italy}

\begin{abstract}
In this work we investigate the relation between curved momentum space and momentum-dependent gauge fields. While the former is a classic idea that has been shown to be tied to minimal-length models, the latter constitutes a relatively recent development in quantum gravity phenomenology. In particular, the gauge principle in momentum space amounts to a modification of the position operator of the form $\hat{X}^\mu\rightarrow\hat{X}^\mu-g A^\mu (\hat{P})$ akin to a gauge-covariant derivative in momentum space according to the minimal coupling prescription. Here, we derive both effects from a Kaluza-Klein reduction of a higher-dimensional geometry exhibiting curvature in momentum space. The interplay of the emerging gauge fields as well as the remaining curved momentum space lead to modifications of the Heisenberg algebra. While the gauge fields imply Moyal-type noncommutativity dependent on the analogue field strength tensor, the dimensionally reduced curved momentum space geometry translates to a Snyder-type noncommutative geometry.
\end{abstract}

\maketitle

\section{Introduction}

The idea of space being endowed with a minimum length or area has captured a lot of attention over the last couple of decades. On the one hand, a popular model of this kind is given by noncommutative geometry of Moyal-type by which we mean rectilinear coordinates obeying the algebra
\begin{equation}
    \label{nc-geo}
    [\hx^i, \hx^j] = i \Theta^{ij},
\end{equation}
where $\Theta^{ij}$ is an anti-symmetric, constant rank-2 tensor, we work in units in which $\hbar=1$ and the Latin indices stand for $d-1$ spatial coordinates (in this paper we concentrate on non-relativistic single-particle quantum mechanics). The remainder of the Heisenberg algebra is assumed to stay unmodified. Following the Schrödinger-Robertson relation, Eq. \eqref{nc-geo} leads to a minimal quantum of area 
\begin{equation}
    \Delta x^i\Delta x^j\geq \frac{1}{2}|\Theta^{ij}|.
\end{equation} 
A review of noncommutative field theory can be found in \cite{Douglas:2001ba}, and interesting applications of noncommutative geometry to the spectrum of the hydrogen atom are considered in \cite{Chaichian:2000si}. Moreover, noncommutative geometry may cure the singularities found in black holes and other solutions in general relativity \cite{Nicolini:2008aj}.

On the other hand, the minimal-length paradigm, one of the oldest concepts in quantum gravity \cite{Bronstein30,Bronstein36,Mead64}, has experienced an significant increase in popularity in recent times. In that vein, the algebra of single-particle observables may be deformed in an alternative way leading to generalized uncertainty principles \cite{Maggiore93c,Kempf94,Kempf96a,Benczik02,Bosso:2020aqm,Bosso:2022vlz,Bosso:2022rue,Bosso:2022ogb} (for a review see \cite{Hossenfelder:2012jw}). These can be derived from algebras of the form 
\begin{equation}
    [\hx^i,\hp_j]=if^i_j(\hp),\hspace{2cm}[\hx^i,\hx^j]=i \theta (\hp)\hat{j}^{ji},\hspace{2cm}[\hp_i,\hp_j]=0,
\end{equation}
with the (possibly modified) generator of rotations $\hat{j}^{ij}=2\hx^{[i}\hp^{j]}$ and the functions of momenta $f^i_j$ and $\theta$ which are related by Jacobi identities (see \cite{Wagner:2021bqz}).

In this paper we search for a single geometric construction leading to either type of deformed Heisenberg algebra. Indeed, there are two concepts which are to be unified:

1. Curved Momentum space \cite{Wagner:2021bqz,Wagner:2022dkc,Wagner:2022rjg,Singh:2021iqa} and

2. Momentum-dependent gauge fields\footnote{This is a genuinely new approach which is not related to the literature on gauge fields \emph{on} curved momentum space (see \eg \cite{Chizhov:1965owb,Romero:2015hwa}). In the present paper, the gauge-fields themselves depend on the momentum of the considered particle. Thus, they are to satisfy momentum-space analogues to Maxwell's equations (or non-abelian generalizations thereof). This has recently been touched upon in \cite{Copinger:2022qzh}.} \cite{Guendelman:2022gue} arising from a momentum-space local $U(1)$ symmetry.

By analogy with the original Kaluza-Klein theory, we show that these two approaches are intimately related because the momentum-gauge fields can be obtained from a curved, higher-dimensional momentum-space geometry via compactification on a circle in momentum space. The Kaluza Klein approach was originally proposed in coordinate space in \cite{Kaluza:1921tu,Klein:1926tv}, and then further developed in several stages by many authors, some of which have been reviewed in   \cite{Appelquist:1987nr}. This includes examples where the Kaluza Klein approach has been generalized to include many extra dimensions that could lead to non-abelian gauge fields. In this context, the off-diagonal components of the metric, \ie the ones mixing an extra dimension with ordinary dimensions, are reinterpreted as gauge fields. Similarly, gauge invariance is reinterpreted as a subgroup of general coordinate invariance. 

Note that, while we sometimes revert to representations throughout this paper to provide intuitive interpretations, the whole formalism is considered on the level of operators and therefore independent of representation.

The paper is structured as follows. In section \ref{sec:cms}, we introduce the notion of curved momentum space and its connection to generalized uncertainty principles. Section \ref{sec:KKR} is devoted to Kaluza-Klein reduction leading to the main result of the paper. Finally, we conclude in section \ref{sec:Con}.

\section{Curved momentum space\label{sec:cms}}

The idea that geometry may depend on directions additionally to positions originally goes back to Riemann's famous habilitation dissertation \cite{Riemann:1854akd}. Mathematically, it was first developed by Finsler \cite{Finsler:1918aps} and Cartan \cite{Cartan:1934las} to be further expanded upon throughout the whole 20th century (for a comprehensive overview consult \cite{Miron:2001oas,Miron:2012pso}).

It was Born who, under the impression of finding the symmetry of quantum mechanics under exchange of positions and momenta ($\hx\longleftrightarrow -\hp$), proposed to let momentum space be curved \cite{Born:1938ajs}. Indeed, it is the curvature of spacetime, and only spacetime, that breaks this symmetry, now dubbed Born reciprocity. This idea sparked a number of further works, namely by Gol'fand \cite{Golfand:1959osp,Golfand:1962lsu,Golfand:1963kas}, Tamm \cite{Tamm:1965lsg,Tamm:1972kie} as well as Batalin and Fradkin \cite{Batalin1989:jas,Batalin1989:lsc} which culminated in the theory of quantum groups by Drinfel'd \cite{Drinfeld:1988afs} and their physical application by Majid \cite{Drinfeld:1988afs,Majid:1988csh,Majid1990:kas,Majid:1991avf}.\footnote{Momentum-dependent gauge fields, in turn, are the Born reciprocals of ordinary gauge fields. Imposing Born reciprocity on the content of the universe (as Born did it for geometry) would thus make their existence a necessity. This will become even clearer in the present work where both the curved momentum space as well as the momentum-dependent gauge fields stem from the same higher, dimensional geometry.}

It has been known for a some time that a number of models studied in Quantum Gravity Phenomenology \cite{Amelino-Camelia:2008aez,Addazi:2021xuf} find their geometric counterpart in curved momentum space. For instance, Lorentz invariance violating theories can be understood geometrically in terms of Finsler or Hamilton geometries \cite{Barcaroli:2015xda,Carmona:2019fwf,Carmona:2021gbg,Gubitosi:2021itz,Relancio:2022mia} (for the mathematical background see \cite{Miron:2012pso}). Similarly, Deformed Special Relativity \cite{Amelino-Camelia:2000stu,Magueijo:2001cr} is set on a de Sitter-shaped momentum space \cite{Kowalski-Glikman:2003qjp,Franchino-Vinas:2022fkh}.

Recently, it has been shown that modified Heisenberg algebras follow a similar pattern \cite{Wagner:2021bqz,Wagner:2022dkc,Wagner:2022rjg,Singh:2021iqa}: Non-relativistic single-particle theories exhibiting a generalized uncertainty principle as well as a noncommutative geometry can be mapped onto theories yielding non-relativistic single-particle dynamics on curved momentum space. Here, the concept of curved momentum space refers to the theory of generalized Hamilton spaces (see \cite{Miron:2012pso,Wagner:2022rjg} for more information). 

This ansatz may as well be applied in reversed order \cite{Wagner:2022rjg} such that a noncommutative space emerges from an underlying momentum space geometry. Along these lines, the kinematics of the non-relativistic single-particle quantum theory at hand are to be described in terms of the canonical positions $\hX^i$ and momenta $\hP_i,$ satisfying the algebra
\begin{align}
    [\hX^i,\hP_j]=i\delta^i_j,&& [\hX^i,\hX^j]=0,&&[\hP_i,\hP_j]=0.\label{HeisAlg}
\end{align}
Given these rather ordinary kinematics, the background geometry is encoded into the dynamics, \ie the Hamiltonian $\hat{H}$ -- a sum of the kinetic energy $K$ the potential $V.$ 

On the one hand, the kinetic part of the Hamiltonian in itself should give rise geodesic motion on the background Hamilton geometry. Recently, it has been shown that this is the case for the geodesic distance from the origin in momentum space $\sigma_P^2$ \cite{Relancio:2020rys}, satisfying the differential equation
\begin{equation}
    g_{ij}(p)\dot{\partial}^i\sigma_P^2\dot{\partial}^j\sigma_P^2=4\sigma_P^2,\label{MomGeoDef}
\end{equation}
with the momentum-dependent metric $g_{ij}.$ Then, the obvious choice, $K=\sigma_P^2/2m,$ with the mass of the given particle $m$, is covariant with respect to passive diffeomorphisms in momentum space as expected from a theory of curved momentum space. Moreover, it reduces to $K=\delta^{ij}P_iP_j/2m,$ \ie the ordinary free-particle Hamiltonian, in the flat limit. 

On the other hand, any covariant central potential\footnote{On the formal level, the family of potentials which can be considered along the lines of  the present paper coincides with the one described in \cite{Gamboa:2000yq}.} depends on the geodesic distance from a point, the origin say, \ie $V(\sigma_X^2).$ As example may serve the isotropic harmonic oscillator of frequency $\omega$, implying $V=m\omega^2\sigma_X^2/2$. The nontrivial step here lies in assuming both geodesic distances, $\sigma_P$ and $\sigma_X,$ to be derived from the \emph{same} metric. This metric thus governs both position and momentum space simultaneously which has been shown to be the case in Hamilton geometries \cite{Miron:2001oas}.\footnote{Beyond noncommutative geometry in quantum gravity phenomenology, it is not uncommon to find Hamiltonians which are not quadratic in momenta (for example in polymer quantum mechanics \cite{Corichi:2006qf,Corichi:2007tf} or the Dicke model \cite{Emary:2003zza}). In principle, depending on the model at hand, it may also be possible to reinterpret these in terms of curved momentum space, as long as the to-become geodesic distances in position and momentum space can indeed be derived from the same metric.}

In a nutshell, the Hamiltonian governing the non-relativistic dynamics on a given curved momentum space reads
\begin{equation}
    \hat{H}=\frac{\hat{\sigma}_P^2}{2m}+V(\hat{\sigma}_X^2),\label{CurvHam}
\end{equation}
where $\hat{\sigma}_P$ and $\hat{\sigma}_X$ represent the operator-valued counterparts of the geodesic distances to be specified below.\footnote{On a curved momentum space, the position representation of these operators may appear to be hopelessly nonlocal. However, for phenomenological purposes, it is usually enough to revert to perturbative techniques. On the nonperturbative level, it is more practical to make use of the momentum representation.}

Given a curved background, commutation relations of the form provided in Eq. \eqref{HeisAlg} are satisfied by phase space variables representing normal coordinates and their conjugates, a fact that was already appreciated by Dirac in 1930 \cite{Dirac30}. Assume, by analogy, the given system to be described in terms of normal coordinates constructed around the origin of momentum space. Then, the classical geodesic distances introduced above read $\sigma_X^2=X^i g_{ij}(P)X^j$ and $\sigma_P^2=\delta^{ij}P_i P_j,$ respectively, with the Kronecker symbol $\delta^{ij}.$ The quantum mechanical counterpart of $\sigma_X^2$ clearly suffers from an operator ordering ambiguity. This ambiguity, however, can be resolved by reverting to geometric calculus as outlined in \cite{Pavsic01} in the context of curved position space. As a result, quantization is carried through in a local orthonormal frame $\hat{x}^a=e^{a}_i(\hP)\hat{X}^i$ as $\hat{\sigma}_X^2=\delta_{ab}\hx^a\hx^b,$ with the \emph{vielbein} $e^a_i.$\footnote{Accordingly, a momentum representation in which $\hat{\sigma}_X^2$ is self-adjoint requires to choose the invariant volume form $\sqrt{\det g^{ij}(P)}\mathrm{d}^{d-1}P$ (in $d-1$ dimensions) as Hilbert space measure.} It is important to note, however, that the ordering is irrelevant for the interpretation of the underlying geometry, which (as usual in first-quantized theories) is intrinsically classical. 

Say, we decide to describe the system in question in terms of the normal-frame positions $\hx^a$ and the original momenta. This amounts to the noncanonical transformation
\begin{equation}
    \hX^i\rightarrow\hx^a=e^a_i\hX^i, \hspace{2cm}\hP_i\rightarrow\hp_a=\delta_a^i\hP_i.\label{trafo}
\end{equation}
These new variables are not of Darboux-type. Instead, they obey the nontrivial algebra
\begin{equation}
    [\hx^a,\hp_b]=ie^a_b,\hspace{1cm} [\hx^a,\hx^b]=2ie^{[a}_i\dot{\partial}^{|i|}(e^{b]}_j)(e^{-1})^j_e\hx^e ,\hspace{1cm} [\hp_a,\hp_b]=0,
\end{equation}
with the derivative in momentum space $\dot{\partial}^j=\partial/\partial P_j=\delta^j_a\partial/\partial p_a.$ Furthermore, the Hamiltonian becomes
\begin{equation}
    H=\frac{\hp^2}{2m}+V(\hx^2),\label{GUPHam}
\end{equation}
modulo additional fields, where we defined $\hx^2\equiv\eta_{ab}\hx^a\hx^b$ and $\hp^2\equiv\eta^{ab}\hp_a\hp_b.$ Thus, we apparently describe dynamics in flat spacetime. This is, of course, not really the case. Instead, we just shifted all geometrical information to the algebra. Then, given a suitably chosen \emph{vielbein}, this construction indeed results in a theory exhibiting a generalized uncertainty principle as well as a noncommutative geometry.\footnote{At this point, it becomes clear why the present discussion is restricted to nonrelativistic systems. While the same noncanonical transformation may also be carried out on the relativistic level in analogous manner, there is no reason to impose that $\hx^2=\sigma_X^2$ because relativistic systems are, in general, not subject to nonlocal potentials of the kind given in Eqs. \eqref{CurvHam} and \eqref{GUPHam}.}

In particular, we may choose to expand the geometry in terms of Riemann normal coordinates around the origin in momentum space (\ie for small momenta) such that the \emph{vielbein} reads approximately
\begin{equation}  \label{ijvielbein}
    e^i_a=\delta^i_a+\frac{1}{6}S_{a}^{~jik}|_{\hP=0}\hP_j\hP_k.
\end{equation}
The resulting modification to the Heisenberg algebra becomes
\begin{align}
	[\hx^a,\hp_b]\simeq & i\left(\delta^a_b+\left.S_b^{~cad}\right|_{p=0}\hp_c\hp_d/6\right),\\
	[\hx^a,\hx^b]\simeq & i\left.\left(S^{bacd}+S^{d[ab]c}\right)\right|_{p=0}\hat{j}_{cd}/6,\\
	[\hp_a,\hp_b]=&0,
\end{align}
with the curvature tensor in momentum space $S^{acbd}.$ For instance a maximally symmetric space satisfying $S^{acbd}=S(g^{ab}g^{cd}-g^{ad}g^{bc})/d(d-1)$ (where $S$ denotes the scalar curvature) implies the deformed algebra
\begin{align}
    [\hx^a,\hp_b]\simeq & i\delta^a_b\left[1+\frac{S\hp^2}{6d(d-1)}\right]-\frac{iS\hp^a\hp_b}{6d(d-1)},\\
	[\hx^a,\hx^b]\simeq & \frac{iS }{2d(d-1)}\hat{j}^{ba},\\
	[\hp_a,\hp_b]=&0.
\end{align}
This surely is of the form contemplated in the literature on generalized uncertainty principles \cite{Maggiore:1993kv,Kempf:1994su,Hossenfelder:2012jw,Bosso:2022rue}. Indeed, it similar but not equivalent to a Snyder space \cite{Snyder46,Franchino-Vinas:2019nqy,Franchino-Vinas:2020umq,Franchino-Vinas:2021bcl}.

\section{Kaluza-Klein reduction and Emergence of momentum gauge fields \label{sec:KKR}}

The basic construction behind Kaluza-Klein theories in position space requires postulating extra dimensions, in the original Kaluza Klein model, one extra dimension, say $d$, in addition to the observed $d-1$ dimensions (indicated by Latin letters, ranging from $1$ until $d-1$ as above). While the metric has $g_{dd}$ and $g_{di}$ components, it is assumed to be independent of the coordinate $X^d.$ This is widely known as the cylinder condition. Then, the gauge fields are identified from the  $g_{di}$ components. 

In the present work, we apply this idea to a momentum-dependent background. By complete analogy, all components of the metric are assumed to be $P_d$-independent, \ie it satisfies the analogue cylinder condition $\partial g_{IJ}/\partial P_d=0.$ The momentum-gauge fields are identified from the $g_{di}$ components.

In that vein, we assume the higher-dimensional background to be described by the metric
\begin{equation}
    g^{IJ}=
    \begin{pmatrix}
    g^{ij}(\hP_k)+\phi^2 A^i A^j(\hP_k)&\phi^2 A^i(\hP_k)\\\phi^2 A^j(\hP_k) &\phi^2(\hP_k)
    \end{pmatrix},\label{FullMet}
\end{equation}
with the dilaton field $\phi$ and the gauge one-form $A_\mu,$ and where the capitalized indices $IJ$ range from $1$ to $d.$ Its inverse can be immediately obtained as
\begin{equation}
    g_{IJ}=
    \begin{pmatrix}
    g_{ij}(\hP_k)&-A^kg_{ki}(\hP_k)\\-A^kg_{kj}(\hP_k) &(A^iA^jg_{ij}-\phi^{-2})(\hP_k)
    \end{pmatrix},\label{FullInvMet}
\end{equation}
with $g_{ij}=(g^{ij})^{-1}.$ Note that this definition is inverse to the usual one in coordinate space -- in momentum space the r\^{o}les of the metric and its inverse are generally interchanged (see \eg \cite{Wagner:2021luc}). 

To obtain the dynamics of a non-relativistic particle moving on this background, we proceed along the lines of the preceding section. However, we have to add the assumption that the particle at hand cannot propagate into the additional, $d$th dimension. Thus, the resulting Hamiltonian has to be evaluated at vanishing $P_d.$ As a result, it has to be of the form
\begin{equation}
    \hat{H}=\frac{\hat{\sigma}_{\hP}^2|_{\hP_d=0}}{2m}+V\left(\hat{\sigma}_{\hX}^2|_{\hP_d=0}\right),
\end{equation}
where $V$ denotes the potential as above. Furthermore, $H\neq H(P_d),$ implies by Noether's theorem that the constant $X^d$ can be interpreted as a charge parameter. As a matter of fact, this is the exact analogue of the coordinate-space Kaluza-Klein theories as a part of which the independence of the metric of the $d$th coordinate leads to the interpretation of $P_d$  as electric charge.

Plugging in the metric \eqref{FullMet}, the geodesic distance from the origin in position space becomes
\begin{equation}
    \sigma_x^2=g_{IJ}X^IX^J=g_{ij}(X^i- X^dA^i)(X^j- X^dA^j)+(X^d)^2/\phi^2.\label{PosGeodDistD}
\end{equation}
This is to be described in terms of $d$ coordinates $x^i$ such that $\sigma_x^2=x^2,$ a procedure that can be simplified by choosing the dilaton to be a large constant. Then, the last term in Eq. \eqref{PosGeodDistD} is negligible. A large interval for the extra dimension in momentum space implies a small interval in coordinate space. Thus, the extra dimension is assumed to be very small as usual in Kaluza-Klein theories. As a result, the inverse metric reads
\begin{equation}
        g_{IJ}=
    \begin{pmatrix}
    g_{ij}&A_i\\A_j &A_k A^k
    \end{pmatrix}.\label{FullMetPlugged}
\end{equation}
Applying this identification, we immediately obtain
\begin{equation}
    \sigma_x^2=g_{ij}(X^i-X^dA^i)(X^j-X^dA^j).
\end{equation}
Along the lines of the preceding section, this quantity may be quantized in a local orthonormal frame. Thus, it may be expressed as 
\begin{equation}
    \hat{\sigma}^2_x=\delta_{ab}\hx^a\hx^b,
\end{equation}
with the new coordinate
\begin{equation}
    \hx^a=e^a_i(\hX^i - X^{d} A^i(\hP_k)).\label{minimakkycoupledcoordinatefromKK}
\end{equation}

%The remaining $ij$-components of the metric can then be expanded according to equation  (\ref{ijvielbein}), where now $S_{j}^{~kjl}|_{\hP_i=0}$ represents the curvature of the $\hP_N = constant$ slices evaluated at zero $\hP_i$. 

Concerning the kinetic term, the projection on $P_d=0,$ which is valid on the whole geodesic, implies that the geodesic distance from the origin in momentum space is evaluated on the corresponding hypersurfaces. By analogy with Eq. \eqref{MomGeoDef}, it then satisfies the differential equation \cite{Synge60}
\begin{equation}
    4\sigma_p^2|_{P_d=0}=g_{IJ}\dot{\partial}^I\sigma_p^2|_{P_d=0}\dot{\partial}^J\sigma_p^2|_{P_d=0}=g_{ij}\dot{\partial}^i\sigma_p^2|_{P_d=0}\dot{\partial}^j\sigma_p^2|_{P_d=0}.
\end{equation}
As a result, the kinetic term is independent of the momentum gauge fields and can be derived solely from the lower-dimensional metric. 

Additionally assuming as in the preceding section that the original system was described in terms of normal coordinates expanded around the origin in momentum space, the geodesic distance can be expressed as $\sigma_p^2=\delta^{ij}P_i P_j.$ In this case, we may thus complement the transformation of the coordinates with an identity transformation of the momenta such that we generalize Eq. \eqref{trafo} to
\begin{equation}
        \hx^a=e^a_i\left[\hX^i-X^d A^i(\hP_k)\right], \hspace{2cm}\hp_a=\delta_a^i\hP_i.\label{fulltrafo}
\end{equation}

Bearing in mind the momentum-space representation of the position operator (${\hat X}^i = i \partial /\partial P_i$), Eq. \eqref{fulltrafo} exactly amounts to the minimal-coupling prescription of the momentum derivative (the position operator) discussed in \cite{Guendelman:2022gue}, \ie a gauge-covariant derivative in momentum space $\dot{\text{D}}_\mu$, where $X^d$ plays the r\^{o}le of a charge. Indeed, it may be understood as a momentum-space local symmetry of the quantum dynamics with respect to $U(1)$-transformations of the wave function \ie $\psi(p)\rightarrow e^{i\alpha(p)}\psi(p).$ Note, however, that while this is an appealing interpretation, on the formal level \emph{this approach is not tied to any representation}. In principle, it is also possible to interpret this symmetry in position space, in the form $\psi(x)\rightarrow e^{i\alpha(-i\partial)}\psi(x).$

Given the transformation \eqref{fulltrafo}, we obtain the general modified algebra
\begin{equation}
    [\hx^a,\hp_b]=ie^a_b,\hspace{1cm} [\hx^a,\hx^b]=2ie^a_i\dot{\partial}^i(e^b_j)(e^{-1})^j_e\hx^e+i(X^d)^2F^{ab}, \hspace{1cm} [\hp_a,\hp_b]=0,
\end{equation}
with the field strength of the momentum gauge field $F^{ab}=2\dot{\partial}^{[a}A^{b]}$ (which was to be expected because $F^{ab}\equiv [\dot{\text{D}}^a,\dot{\text{D}}^b]/(X^d)^2$). For constant field strength $(X^d)^2F^{ab}=\Theta^{ab},$ \ie to lowest admissible order, the noncommutativity induced by the momentum gauge fields is of Moyal-type as provided in Eq. \eqref{nc-geo}.

To subleading order of the Riemann normal coordinate expansion on maximally symmetric slices, this combines Moyal- with Snyder-type noncommutativity, reading
\begin{align}
    [\hx^a,\hp_b]\simeq & i\delta^a_b\left[1+\frac{S\hp^2}{6d(d-1)}\right]-\frac{S\hp^a\hp_b}{6d(d-1)},\\
	[\hx^a,\hx^b]\simeq & i\Theta^{ab}+\frac{i S}{2d(d-1)}\hat{j}^{ba},\\
	[\hp_a,\hp_b]=&0.
\end{align}
Thus, both kinds of modification can be derived within the same framework from a curved higher-dimensional geometry in momentum space.

\section{Conclusion\label{sec:Con}}

In this paper we have studied the remarkable connection between two apparently different approaches to the origin of spacetime noncommutativity. The first of these is based on curved momentum space \cite{Wagner:2021bqz,Wagner:2022dkc,Wagner:2022rjg,Singh:2021iqa}, while the second one involves momentum-space analogues of the well-known $U(1)$ gauge fields \cite{Guendelman:2022gue} expressing an invariance under momentum-space local $U(1)$ transformations. Both of these formulations combine an unusual Hamiltonian (dynamics) with an ordinary algebra of observables (kinematics). We find equivalent descriptions for these theories in terms of an ordinary-looking Hamiltonian, but a nontrivial algebra of observables, thus shifting the deformation from geometry to algebra. In this way, we make contact with the kinds of deformations which are routinely studied in quantum gravity phenomenology.

Drawing on the formalism introduced in \cite{Wagner:2022dkc}, by means of a noncanonical transformation of the momentum coordinates,  we have converted the Hamiltonian of a single quantum particle moving on a curved momentum space into its flat-space counterpart, while simultaneously obtaining a noncommutative geometry of Snyder-type for the new spacetime coordinates. Moreover, along the lines of \cite{Guendelman:2022gue}, we have obtained coordinate noncommutativity of Moyal-type from the commutator of momentum-space covariant derivatives induced by underlying momentum-dependent gauge fields \cite{Guendelman:2022gue}.

Finally, we have unified both of these approaches in a Kaluza-Klein scenario in momentum space. In that vein, we have added an additional dimension to the background and equipped the underlying metric with non-vanishing off-diagonal elements. Under the assumption, that the additional dimension is small and that the described particle cannot propagate in it, we have found that these off-diagonal terms exactly provide the momentum gauge fields defined in \cite{Guendelman:2022gue}, while the remaining part of the geometry translates to generalized uncertainty principles as pointed out in  \cite{Wagner:2021bqz,Wagner:2022dkc,Wagner:2022rjg,Singh:2021iqa}, both within one and the same formalism. Together they imply Moyal- as well as Snyder-type noncommutativity. 

While the present paper deals with these noncommutative geometries for commutators of positions, by virtue of the Born reciprocal property of quantum theory \cite{Born:1938ajs} the present discussion can be readily extended to noncommutative momentum space in terms of curved position space and ordinary gauge fields. Indeed, at least on the formal level both can be described simultaneously by reverting to canonical phase space variables, while making use of a deformed Hamiltonian. 

Besides from providing a whole new way of understanding noncommutative geometry, this framework may also be of practical benefit -- if a calculation on one side of the duality constructed in this paper, \eg Moyal- or Snyder-noncommutative geometry, turns out to be too involved, it may be useful to consider the other side, \ie momentum-gauge fields or curved momentum space. In particular, the formalism introduced here makes it possible to consider Moyal- and Snyder-noncommutative geometry in a unified way. Furthermore, the new approach of theories invariant under momentum-space local transformations opens up its very own field of possibilities. We hope to report back on this subject in future work.

\section{Acknowledgements}
We would like to thank COST ACTION CA18108 - Quantum gravity phenomenology in the multi-messenger approach (QG-MM) for financial support of our research, in particular for the opportunity the action gave us to meet at the third annual conference in Naples Italy, and start a collaboration that lead to this paper. EG also wants to thank  FQXi
for finantial support. FW acknowledges financial support by the Polish National Research and Development Center (NCBR) project ''UNIWERSYTET 2.0. --  STREFA KARIERY'', POWR.03.05.00-00-Z064/17-00 (2018-2022).

%\appendix

%\section{Generalized Hamilton spaces}

%Curved momentum space can be described covariantly in terms of generalized Hamilton spaces \cite{Miron:2001oas,Miron:2012pso}. 

\bibliographystyle{utphys}
\bibliography{ref.bib}

\begin{comment}

\end{comment}

\end{document}